\documentclass{jpsj3}
\usepackage{txfonts}

\newcommand{\beq}{\begin{equation}}
\newcommand{\eeq}{\end{equation}}
\newcommand{\beqa}{\begin{eqnarray}}
\newcommand{\eeqa}{\end{eqnarray}}
\newcommand{\bit}{\begin{itemize}}
\newcommand{\eit}{\end{itemize}}
\newcommand{\bdes}{\begin{description}}
\newcommand{\edes}{\end{description}}
\newcommand{\bfig}{\begin{figure}}
\newcommand{\efig}{\end{figure}}

\title{The Exact Susceptibility of the Spin-S Transverse Ising Chain with Next-nearest-neighbor Interactions.}

\author{Kazuhiko Minami}
\inst{Graduate School of Mathematics, Nagoya University, \\
Furo-cho, Chikusa-ku, Nagoya, Aichi, 464-8602, Japan }

\abst{
The zero-field susceptibility 
of the spin-$S$ transverse Ising chain 
with next-nearest-neighbor interactions 
is obtained exactly. 
The susceptibility is given in an explicit form for $S=1/2$, 
and expressed in terms of the eigenvectors of the transfer matrix 
for general spin $S$. 
It is found that the low-temperature limit is independent of spin $S$, 
and is divergent at the transition point. 
}

\begin{document}
\maketitle

\section{Introduction}

The transverse Ising model is a simple and basic spin model 
with non-trivial quantum effects. 
Specifically the spin $1/2$ transverse Ising chain has been 
investigated by many authors. 
Fisher\cite{60Fisher}\cite{63Fisher} 
derived the exact zero-field transverse susceptibility, 
Katsura\cite{62Katsura} obtained the free energy, 
and Pfeuty\cite{70Pfeuty} also obtained the free energy 
through the Jordan-Wigner transformation. 
The one-dimensional spin $1/2$ transverse Ising model 
is known to be equivalent 
to the two-dimensional rectangular Ising model. 
\cite{71Suzuki} %\cite{14Minami} 
The Jordan-Wigner transformation was generalized,\cite{16Minami}
and an infinite number of systems 
which can be regarded as kind of generalizations of the transverse Ising chain 
were solved exactly.\cite{16Minami}\cite{17Minami}\cite{20Yanagihara}
Applications of the Jordan-Wigner transformation 
to higher dimensions 
have also been considered by many authors.\cite{07Feng}\cite{07ChenHu}\cite{08ChenNussinov}\cite{19Minami}\cite{22Cao}\cite{22Li}

The Ising chain 
with general spin $S$ have also been investigated. 
\cite{67Suzuki}\cite{68Obokata}\cite{69Dobson}
The zero-field susceptibility of spin $S$ transverse Ising chain 
was exactly derived.\cite{83Thorpe}\cite{85Chatterjee}\cite{96Minami} 
The transverse susceptibility and the specific heat were shown up to $S=60$ 
and the crossover from the $S=1/2$ case to the classical case 
was exactly observed in Ref.\citen{98Minami}.
The transverse susceptibility with a non-zero longitudinal external field, 
and that of Ising models 
with alternate or random structures were also obtained.\cite{98Minami}\cite{13Minami}
The continuous spin chain which can be regarded as the infinite spin limit 
of the Ising chain was also investigated.
\cite{67Joyce}\cite{68Thompson}\cite{90Horiguchi} 

The Ising chain with next-nearest-neighbor interactions 
have also been investigated by many authors. 
The free energy and correlation functions were, 
with the use of a dual transformation, 
derived by Stephenson\cite{70Stephenson} 
and Hornreich et al.\cite{79Hornreich}
The zero-field transverse susceptibility for the spin $1/2$ case 
was calculated by Harada.\cite{83Harada}
Oguchi introduced\cite{65Oguchi} a transfer matrix 
for the Ising chain with nearest- and next-nearest-neighbor interactions, 
and the free energy was re-derived by Kassan-Ogly.\cite{01Kassan-Ogly}

In this paper, 
the exact zero-field susceptibility 
of the transverse Ising chain with the next-nearest-neighbor interactions 
is derived for general spin $S$. 
In section 2.1, 
the transverse susceptibility is expressed as an infinite sum of correlation functions. 
In section 2.2, 
the transfer matrix $V$ is introduced for the Ising chain with the next-nearest-neighbor interactions. 
The transfer matrix for the $S=1/2$ case is explicitly considered in detail as an example. 
In section 2.3, 
the susceptibility is expressed as a finite sum  
using the eigenvectors of $V$ for general spin $S$, 
and written down explicitly specifically for $S=1/2$ . 
In section 3, 
the low-temperature limit is considered, 
and it is derived that the $T\to 0$ limit is independent of $S$, 
and is divergent at the point where the ground state shows phase transition.

\section{Formulation}

\subsection{Model and Formulation}

%------　模型の定義
%

Let us consider the Hamiltonian  
\beq
{\cal H}=({\cal H}_{0}+{\cal H}_{1})-H{\cal Q},
\label{ham}
\eeq
where 
\begin{eqnarray}
{\cal H}_{0}
=
-J_{0}\sum_{j=1}^{N}s_{j}^zs_{j+1}^z,  
\hspace{0.8cm}
{\cal H}_{1}
=
-J_{1}\sum_{j=1}^{N}s_{j}^zs_{j+2}^z, 
\hspace{0.8cm}  
{\cal Q}  
=g\mu_{\rm B}\sum_{j=1}^{N}s_{j}^x,
\end{eqnarray}
and the periodic boundary condition is assumed. 
Here ${\cal H}_{\rm Is}={\cal H}_{0}+{\cal H}_{1}$ 
is the Ising interactions, 
and ${\cal Q}$ is the transverse term 
which does not commute with ${\cal H}_{\rm Is}$.

%------　帯磁率の一般式
%

The transverse susceptibility $\chi$ at zero-field is defined as 
\beqa
\chi&=&\frac{\partial}{\partial H}\langle {\cal Q} \rangle |_{H=0}
\nonumber\\
    &=&\frac{\partial}{\partial H}
       \frac{{\rm Tr\:} {\cal Q} \exp(-\beta{\cal H})}
            {{\rm Tr} \exp(-\beta{\cal H})}\Big|_{H=0}.
\label{susdef}
\eeqa
The expectation $\langle\:\rangle$ is taken by ${\cal H}$, 
and $\beta=1/k_{\rm B}T$ 
where $k_{\rm B}$ is the Boltzmann constant 
and $T$ is the temperature.  
For the purpose to calculate (\ref{susdef}), 
let us introduce the formula 
\beq
\frac{d}{dH}\exp(-\beta{\cal H})=
\exp(-\beta{\cal H})
\int_0^\beta \exp(\lambda{\cal H}){\cal Q}\exp(-\lambda{\cal H})d\lambda, 
\label{opder}
\eeq
which can be derived by multiplying $\exp(\beta{\cal H})$ 
from the left-hand side and differentiating in terms of $\beta$. 
Using (\ref{opder}), 
$\langle {\cal Q}\rangle$ 
can be expanded in terms of $H$, and we obtain 
\beqa
\chi
&=&
\frac{\partial}{\partial H}
      [\frac{\langle{\cal Q}\rangle_0
             +H\int_0^\beta\langle\exp(\lambda{\cal H}_{\rm Is}){\cal Q}
               \exp(-\lambda{\cal H}_{\rm Is}){\cal Q}\rangle_0 d\lambda}
             {1+H\int_0^\beta\langle{\cal Q}\rangle_0 d\lambda}+O(H^2)]\Big|_{H=0}
\nonumber\\
&=&
\int_0^\beta 
\langle
\exp(\lambda{\cal H}_{\rm Is}){\cal Q}\exp(-\lambda{\cal H}_{\rm Is}){\cal Q}
\rangle_0 d\lambda,
\label{sus-1}
\eeqa
where $\langle\:\rangle_0$ is the expectation 
taken at zero external field $H=0$. 

Next let us introduce inner derivatives $\delta_{0}$ and $\delta_{1}$
through the relations 
\begin{eqnarray}
\delta_{0}^0{\cal Q}={\cal Q}, \hspace{0.8cm} 
\delta_{0}^{n+1}{\cal Q}=[{\cal H}_{0},\delta_{0}^n{\cal Q}],
\nonumber\\
\delta_{1}^0{\cal Q}={\cal Q}, \hspace{0.8cm} 
\delta_{1}^{n+1}{\cal Q}=[{\cal H}_{1},\delta_{1}^n{\cal Q}], 
\end{eqnarray}
and introduce $\delta=\delta_{0}+\delta_{1}$. 
The expectation in (\ref{sus-1}) can be expanded using 
\beq
\exp(\lambda{\cal H}_{\rm Is}){\cal Q}\exp(-\lambda{\cal H}_{\rm Is})=
\sum_{p=0}^\infty\frac{\lambda^p}{p!}\delta^p{\cal Q},
\eeq
and (\ref{sus-1}) is expressed as
\beq
\chi=\int_0^\beta 
     \langle
     \sum_{p=0}^\infty\frac{\lambda^p}{p!}\delta^p{\cal Q}\:{\cal Q}
     \rangle_0 d\lambda.
     \label{chiint}
\eeq
It can be seen that (\ref{chiint}) is equal to 
\beqa
\chi&=&\sum_{p=0}^\infty\frac{1}{p!}
       \int_0^\beta \lambda^p d\lambda
       \langle\delta^p{\cal Q}\:{\cal Q}\rangle_0
\nonumber\\
    &=&\sum_{p=0}^\infty\frac{\beta^{p+1}}{(p+1)!}
       \langle\delta^p{\cal Q}\:{\cal Q}\rangle_0.
\label{sus-2}
\eeqa
This formula was already derived in Ref.\citen{96Minami}. 
What we have to do next is to evaluate the correlation functions 
$\langle\delta^p{\cal Q}\:{\cal Q}\rangle_0$ 
with non-zero next-nearest-neighbor interaction, 
and to perform the infinite sum in (\ref{sus-2}).

%------　相関関数の展開
% 7/21/2022-6から7/23/2022-1

Then let us consider $\langle\delta^p{\cal Q}\:{\cal Q}\rangle_0$. 
Because of the commutatibity 
${\cal H}_{1}{\cal H}_{0}={\cal H}_{0}{\cal H}_{1}$, 
we generally find 
$\delta_{1}\delta_{0}=\delta_{0}\delta_{1}$,  
and then $\delta^p{\cal Q}$ is written as 
\begin{eqnarray}
\delta^p{\cal Q}
&=&
(\delta_{0}+\delta_{1})^{p}{\cal Q}
=
g\mu_{\rm B}
\sum_{l=0}^{p}
\left(
\begin{array}{c}
p\\
l
\end{array}
\right)
\delta_{0}^{p-l}\delta_{1}^{l}\sum_{j=1}^{N}s^{x}_{j},
\label{deltapQ}
\end{eqnarray}
where 
\begin{eqnarray}
\left(\begin{array}{c}p\\l\end{array}\right)=
\frac{p!}{(p-l)!l!}.
\end{eqnarray}
%--- see 7/22/2022-3、7/23/2022-1
It is straightforward to show inductively that 
\begin{eqnarray}
&&
\delta_{0}^{p_{0}}\delta_{1}^{p_{1}}
\sum_{j=1}^{N}s^{x}_{j}
\nonumber\\
&&
=
\sum_{j=1}^{N}
\Big(
(-J_{0})^{p_{0}}
\sum_{l_{0}=0}^{p_{0}}
\left(
\begin{array}{c}
p_{0}\\
l_{0}
\end{array}
\right)
(s^{z}_{j-1})^{p_{0}-l_{0}}(s^{z}_{j+1})^{l_{0}}
\Big)
\Big(
(-J_{1})^{p_{1}}\sum_{l_{1}=0}^{p_{1}}
\left(
\begin{array}{c}
p_{1}\\
l_{1}
\end{array}
\right)
(s^{z}_{j-2})^{p_{1}-l_{1}}(s^{z}_{j+2})^{l_{1}}
\Big)
\tau_{p_{0}+p_{1}}s^{k(p_{0}+p_{1})}_{j},
\nonumber\\
\label{delta1delta0}
\end{eqnarray}
where 
\begin{eqnarray}
\tau_{p}
=
\left\{
\begin{array}{cl}
1&\hspace{0.3cm}p={\rm even}\\
i&\hspace{0.3cm}p={\rm odd}
\end{array}
\right.
\end{eqnarray}
and $k(p)=x$ for even $p$, and $k(p)=y$ for odd $p$. 
Then from (\ref{deltapQ}) and (\ref{delta1delta0}), we find 
\begin{eqnarray}
\langle\delta^p{\cal Q}\:{\cal Q}\rangle_0
&=&
(g\mu_{\rm B})^{2}
\sum_{l=0}^{p}
\left(
\begin{array}{c}
p\\
l
\end{array}
\right)
\langle
\delta_{0}^{p-l}\delta_{1}^{l}
\sum_{j=1}^{N}s^{x}_{j}, 
\sum_{j'=1}^{N}s^{x}_{j'}
\rangle_0
\nonumber\\
&=&
(g\mu_{\rm B})^{2}
\sum_{j=1}^{N}
\sum_{l=0}^{p}
\left(
\begin{array}{c}
p\\
l
\end{array}
\right)
(-J_{0})^{p-l}(-J_{1})^{l}
\nonumber\\
&\times&
\sum_{l_0=0}^{p_{0}}
\left(
\begin{array}{c} 
p_{0}\\ 
l_0 
\end{array}
\right)
\sum_{l_1=0}^{p_{1}}
\left(
\begin{array}{c} 
p_{1}\\ 
l_1 
\end{array}
\right)
\langle
(s^{z}_{j-2})^{p_{1}-l_{1}}
(s^{z}_{j-1})^{p_{0}-l_{0}}
(\tau_{p}s^{k(p)}_{j}s^{x}_{j})
(s^{z}_{j+1})^{l_{0}}
(s^{z}_{j+2})^{l_{1}}
\rangle_0,
\nonumber\\
\label{deltapQQ}
\end{eqnarray}
where $p_{0}=p-l$, $p_{1}=l$ and thus $p_{0}+p_{1}=p$. 
Here we used the fact that eigenstates of ${\cal H}_{\rm Is}$ 
are direct products of eigenstates of each $s_j^z$, 
and hence 
$\langle
(s^{z}_{j-2})^{p_{1}-l_{1}}
(s^{z}_{j-1})^{p_{0}-l_{0}}
\tau_{p}s^{k(p)}_{j}
(s^{z}_{j+1})^{l_{0}}
(s^{z}_{j+2})^{l_{1}}
\cdot
s^{x}_{j'}
\rangle_0
=0$ 
when $j'\ne j$. 
The correlation function in (\ref{deltapQQ}) can be obtained 
through the transfer matrix method.

\subsection{Transfer Matrix}

%------　伝送行列
% 7/28/2022-1修正から7/28/2022-27修正

The transfer matrix is defined as the matrix 
composed of the Boltzmann factors between two sites   
as its elements: 
\begin{eqnarray}
V=\sum_{s_{j}, s_{j+1}} |s_{j+1}\rangle e^{Ks_{j}s_{j+1}}\langle s_{j}|, 
\end{eqnarray}
and thus 
\begin{eqnarray}
\langle s_{j+1}|V|s_{j}\rangle=e^{Ks_{j}s_{j+1}}. 
\end{eqnarray}
Here we consider the state 
$|s_{j}\rangle_{j}$ at site $j$ 
with the eigenvalue $s_{j}$ $(s_{j}=S, S-1, \ldots, -S)$: 
$s^{z}_{j}|s_{j}\rangle_{j}=s_{j}|s_{j}\rangle_{j}$ , 
and $|s_{j}\rangle_{j}$ is written $|s_{j}\rangle$ for abbreviation.
If one operates $V$ from the left-hand side, 
then one additional bond is introduced to the chain; 
for example $V|s_{j}=s\rangle$ 
provides all the $(j+1)$-th states $|s_{j+1}\rangle$ 
multiplied by the Boltzmann factor obtained under the condition 
that the site $j$ has fixed eigenvalue $s_{j}=s$. 
In our case, 
we consider the configurations of adjacent two sites 
to introduce the next-nearest-neighbor interactions. 
The transfer matrix, in our notation, is defined as 
\begin{eqnarray}
V=\sum_{s_{j-1}, s_{j},s'_{j}, s'_{j+1}}
 |s'_{j}\rangle|s'_{j+1}\rangle 
e^{K_0s_{j-1}s'_{j}}
e^{K_1s_{j-1}s'_{j+1}}
\delta_{s_{j}s'_{j}}
\langle s_{j-1}|\langle s_{j}|, 
\label{trmatgen}
\end{eqnarray}
and thus 
\begin{eqnarray}
\langle s'_{j}|\langle s'_{j+1}|\:V\:|s_{j-1}\rangle|s_{j}\rangle
=
e^{K_0s_{j-1}s'_{j}}
e^{K_1s_{j-1}s'_{j+1}}
\delta_{s_{j}s'_{j}},
\end{eqnarray}
where 
$K_{0}=\beta J_{0}$, $K_{1}=\beta J_{1}$,
and $s_{j-1}, s_{j},s'_{j}, s'_{j+1}=S, S-1, \ldots, -S$.

% 7/28/2022-9
For example in the case of $S=1/2$, we find  
\begin{eqnarray}
V=
\left(
\begin{array}{cccc}
xy & 0 & 1/xy & 0 \\
x/y & 0 & y/x & 0 \\
0 & y/x & 0 & x/y \\
0 & 1/xy & 0 & xy 
\end{array}
\right)
\label{Vseq12}
\end{eqnarray}
where 
$x=e^{\frac{1}{4}K_0}$ and $y=e^{\frac{1}{4}K_1}$. 
The configurations $\{|s_{j-1}\rangle|s_{j}\rangle\}$ 
and $\{|s'_{j}\rangle|s'_{j+1}\rangle\}$
are both arranged by binary rules as 
$
|\frac{1}{2}\rangle|\frac{1}{2}\rangle, 
|\frac{1}{2}\rangle|\frac{-1}{2}\rangle, 
|\frac{-1}{2}\rangle|\frac{1}{2}\rangle, 
|\frac{-1}{2}\rangle|\frac{-1}{2}\rangle 
$,
in which each configuration corresponds to the 
$\rho=1, 2, 3, 4$-th element of the matrix, respectively. 
When one operate the transfer matrix from the left-hand side, 
then the Boltzmann weights coming from an additional site 
are introduced. 
Configurations with the inconsistency $s_{j}\neq s^{'}_{j}$ are eliminated  
multiplying $0$ element. 
Then we find the matrix elements $V_{\rho\rho'}$ for example 
$(V)_{11}
=\langle \frac{1}{2}|\langle \frac{1}{2}|\:V\:|\frac{1}{2}\rangle|\frac{1}{2}\rangle
=xy$, 
$(V)_{21}
=\langle \frac{1}{2}|\langle \frac{-1}{2}|\:V\:|\frac{1}{2}\rangle|\frac{1}{2}\rangle
=x/y$, 
and 
$(V)_{12}
=\langle \frac{1}{2}|\langle \frac{1}{2}|\:V\:|\frac{1}{2}\rangle|\frac{-1}{2}\rangle
=0$. 
The eigenequation is 
\begin{eqnarray}
0
&=&
{\rm det\:}(\lambda E -V)
\nonumber\\
&=&
\lambda^4-2xy\lambda^3+y^2(x^2-\frac{1}{x^2})\lambda^2
+2\frac{y}{x}(y^2-\frac{1}{y^2})\lambda -(y^2-\frac{1}{y^2})^2
\nonumber\\
&=&
\Big[
\lambda^2-y(x+\frac{1}{x})\lambda+(y^2-\frac{1}{y^2})
\Big]
\Big[
\lambda^2-y(x-\frac{1}{x})\lambda-(y^2-\frac{1}{y^2})
\Big].
\end{eqnarray}
The eigenvalues are obtained as 
\begin{eqnarray}
\lambda_1^\pm
&=&
\frac{1}{2}
\Big[
y(x+\frac{1}{x})\pm \sqrt{y^2(x+\frac{1}{x})^2-4(y^2-\frac{1}{y^2})}
\Big],
\end{eqnarray}
and
\begin{eqnarray}
\lambda_2^\pm
&=&
\frac{1}{2}
\Big[
y(x-\frac{1}{x})\pm \sqrt{y^2(x-\frac{1}{x})^2+4(y^2-\frac{1}{y^2})}
\Big].
\end{eqnarray}
When $J_1=0$, then $y=1$, 
and we find 
$\displaystyle \lambda_1^+=x+\frac{1}{x}$, $\lambda_1^-=0$, 
$\displaystyle \lambda_2^+=x-\frac{1}{x}$, and $\lambda_2^-=0$. 
When $J_0=0$, then $x=1$, 
and we find 
$\displaystyle \lambda_1^+=y+\frac{1}{y}$, 
$\displaystyle \lambda_1^-=y-\frac{1}{y}$, 
and $\lambda_2^\pm=\pm\sqrt{\displaystyle y^2-\frac{1}{y^{2}}}$. 
The eigenvector 
corresponding to $\lambda_1^+$ is obtained as 
\begin{eqnarray}
{\bf u}_1
=
\left(
\begin{array}{c}
 1/xy  \\
 u_{-} \\
 u_{-} \\
 1/xy  
\end{array}
\right),
\end{eqnarray}
where
\begin{eqnarray}
u_{\pm}&=&\frac{1}{2} \Big(R_- \pm y\:(x-\frac{1}{x})\Big),
\nonumber\\
R_-&=&\sqrt{y^2(x+\frac{1}{x})^2-4(y^2-\frac{1}{y^{2}})}.
\label{uRpm}
\end{eqnarray}
%The elements of ${\bf u}_{1}$ are all positive, hence from the Perron-Frobenius theorem, $\lambda_1^+$  is the maximum eigenvalue for all $J_0$ and $J_1$. 
The elements of $V$ are all non-negative, and $V$ is irreducible. 
Hence from the Perron-Frobenius theorem, 
the maximum eigenvalue of $V$ is unique. 
Together with the fact that $\lambda_1^+$ is the maximum eigenvalue 
for $J_1=0$ and for $J_0=0$, 
$\lambda_1^+$ is also the maximum for all $J_0$ and $J_1$. 
Let $U$ be a matrix whose column vectors are the eigenvectors of $V$, 
the first column being the eigenvector ${\bf u}_1$. 
Then the first row vector ${\bar{\bf u}}_1$ of the matrix $U^{-1}$ 
is obtained as 
\begin{eqnarray}
{\bar{\bf u}}_1
=
\frac{xy}{2R_-}
\Big( u_{+}, \frac{1}{xy},\frac{1}{xy}, u_{+}\Big).
\end{eqnarray}
It is easy to confirm that ${\bar{\bf u}}_1\cdot{\bf u}_1=1$.

\subsection{The Transverse Susceptibility}

%---帯磁率の計算
%一般式8/2/2022-1、8/28/2022-1

Let us then consider the expectation value in (\ref{deltapQQ}) 
for general spin $S$. 
The quantum expectation of this type can be calculated 
using the transfer matrix method.\cite{96Minami}
Let $\{|s_{1},\ldots, s_{N}\rangle\}$ 
be a complete set of the eigenstates of ${\cal H}_{\rm Is}$, 
in which each 
$|s_{1},\ldots, s_{N}\rangle=|s_{1}\rangle|s_{2}\rangle\cdots|s_{N}\rangle$ 
is a direct product of eigenstates of $s_j^z$ $(j=1, \ldots, N)$.  
Then, the expectation value $\langle s_j^k s_j^x\rangle_0 $, for example, 
can be calculated using the transfer matrix $V$ as 
\beqa
\langle s_j^k s_j^x\rangle_0 &=&
\frac{{\rm Tr\:} s_j^k s_j^x\exp(-\beta{\cal H}_{\rm Is})}{{\rm Tr\:}\exp(-\beta{\cal H}_{\rm Is})} 
=\frac{\displaystyle \sum_{\{s_{j}\}}\langle s_{1},\ldots, s_{N}| s_j^k s_j^x \exp(-\beta{\cal H}_{\rm Is})|s_{1},\ldots, s_{N}\rangle}
{\displaystyle \sum_{\{s_{j}\}}\langle s_{1},\ldots, s_{N}| \exp(-\beta{\cal H}_{\rm Is})|s_{1},\ldots, s_{N}\rangle}
\nonumber\\
&=&\frac{{\rm Tr\:} D_{j}^{(p)}V^N}{{\rm Tr\:} V^N},
\eeqa
where 
$D_{j}^{(p)}$ is an operator defined by 
\begin{eqnarray}
D_{j}^{(p)}=\sum_{s'_{j}, s'_{j+1}} 
|s'_{j}\rangle |s'_{j+1}\rangle \Delta_\rho^{(p)}\langle s'_{j}| \langle s'_{j+1}|, 
\end{eqnarray}
and represented by a diagonal matrix $D^{(p)}$ with the elements 
\beqa
(D^{(p)})_{\rho \rho'}&=&\Delta_\rho^{(p)}\delta_{\rho\rho'},
\nonumber\\
\Delta_\rho^{(p)}&=&
\left\{\begin{array}{ll}
       (s^x s^x)_{ll}=(S(S+1)-(S-l+1)^2)/2&p={\rm even}\\
       (s^y s^x)_{ll}=(S-l+1)/2i& p={\rm odd}, 
        \end{array}
\right.
\eeqa
and $\rho$ is the index 
which corresponds to the spin configuration $(s'_{j}, s'_{j+1})$ 
of the adjacent two sites $j$ and $j+1$ 
(see for example below (\ref{Vseq12})), 
and $l$ is defined by $s'_{j}=S-l+1$, 
i.e. $\rho$ denotes a configuration $(s'_{j}, s'_{j+1})$, 
and $l$ is determined from $s'_{j}$. 
In the case of $S=1/2$, for example, 
$D^{(p)}$ is a diagonal matrix 
$D^{(p)}={\rm diag\:}(1/4, 1/4, 1/4, 1/4)$ when $p$ is even, 
and 
$iD^{(p)}={\rm diag\:}(1/4, 1/4, -1/4, -1/4)$ when $p$ is odd. 
Let us introduce a parameter $m_\rho=s'_{j}=S-l+1$ 
which is the eigenvalue 
satisfying $s_{j}^z|s'_{j}\rangle=m_\rho|s'_{j}\rangle$. 
The parameter $m_\rho$ will be used below.

Similarly the correlation function in (\ref{deltapQQ}) 
for general spin $S$ 
is then expressed, using the transfer matrix (\ref{trmatgen}) 
and the operator 
\begin{eqnarray}
M_{j}=\sum_{s'_{j}, s'_{j+1}} 
|s'_{j}\rangle |s'_{j+1}\rangle s'_{j}\langle s'_{j}| \langle s'_{j+1}|, 
\end{eqnarray}
which is represented by a diagonal matrix 
$M={\rm diag}(m_{1}, \ldots, m_{n})$, 
as 
\begin{eqnarray}
&&
\langle 
(s_{j-2}^z)^{p_{1}-l_1} (s_{j-1}^z)^{p_{0}-l_0}
(\tau_ps_{j}^{k(p)} s_{j}^x)
(s_{j+1}^z)^{l_0} (s_{j+2}^z)^{l_1}
\rangle_0
=
\frac{{\rm Tr\:}M^{p_{1}-l_1} VM^{p_{0}-l_0} V\tau_pD^{(p)} VM^{l_0}VM^{l_1}V^{N-4}}{{\rm Tr\:}V^N}
\nonumber\\
&=&
\frac{
{\rm Tr\:}M^{p_{1}-l_1} VM^{p_{0}-l_0} V\tau_pD^{(p)} VM^{l_0}VM^{l_1}
U
\left(
\begin{array}{cccc}
1&&&\\
&(\lambda_2/\lambda_1)^{N-4}&&\\
&&\ddots &\\
&&&(\lambda_n/\lambda_1)^{N-4}
\end{array}\right)
U^{-1}
}
{
{\rm Tr\:}V^4
U
\left(
\begin{array}{cccc}
1&&&\\
&(\lambda_2/\lambda_1)^{N-4}&&\\
&&\ddots &\\
&&&(\lambda_n/\lambda_1)^{N-4}
\end{array}\right)
U^{-1}
},
\label{corr4}
\end{eqnarray}
where $U$ is a matrix whose column vectors are the eigenvectors of $V$, 
the first column being the eigenvector 
that corresponds to the unique maximum eigenvalue $\lambda_{1}$, 
and $n=(2S+1)^{2}$. 
Matrix $U^{-1}VU=\Lambda$ is a diagonal matrix 
whose diagonal elements are the eigenvalues of $V$. 
Here let us consider the thermodynamic limit $N\rightarrow\infty$, 
then we find 
$(\lambda_{\rho}/\lambda_{1})^{N-4}\to 0$ for $\rho\neq 1$, 
and (\ref{corr4}) is expressed as
\beq
\rightarrow
\frac{1}{c}
\sum_{\iota,\:\kappa=1}^n 
\sum_{\mu,\:\rho,\:\nu=1}^n 
{\bar u}_{1\iota}
m_\iota^{p_{1}-l_1}V_{\iota\mu}m_\mu^{p_{0}-l_0}V_{\mu\rho}
\tau_p \Delta_\rho^{(p)}
V_{\rho\nu}m_\nu^{l_0}V_{\nu\kappa}m_{\kappa}^{l_1}
u_{\kappa 1}
\hspace{0.6cm}
(N\to\infty),
\label{deltapQQlim}
\eeq
where 
\beqa
c
=
\sum_{\iota,\:\kappa=1}^n 
\sum_{\mu,\:\rho,\:\nu=1}^n 
{\bar u}_{1\iota}
V_{\iota\mu}V_{\mu\rho}V_{\rho\nu}V_{\nu\kappa}
u_{\kappa1}
=
\lambda_{1}^{4},
\label{cgenform}
\eeqa
and ${\bar u}_{1\iota}=(U^{-1})_{1\iota}$, $u_{\kappa 1}=(U)_{\kappa 1}$. 
The factor 
${\bar u}_{1\iota}u_{\kappa 1}$ is the exactly renormalized Boltzmann weight 
with fixed configurations $\iota$ and $\kappa$. 
%7/29/2022-3
When the transfer matrix $V$ cannot be diagonalized, 
which occur for example 
when $y^{2}(x+1/x)/2=1$ with $x\neq 1$ in the case of spin $1/2$, 
we can consider the Jordan normal form, 
and obtain an identical result. 
The susceptibility in the thermodynamic limit  is obtained, 
from (\ref{sus-2}), (\ref{deltapQQ}) and (\ref{deltapQQlim}), as 
\beqa
\chi_0
&=&\lim_{N\rightarrow\infty}\frac{\chi}{N}
\nonumber\\
&=&\sum_{p=0}^\infty\frac{\beta^{p+1}}{(p+1)!}
   \lim_{N\rightarrow\infty}\frac{1}{N}\langle\delta^p{\cal Q}{\cal Q}\rangle_0
\nonumber\\
&=&
\sum_{p=0}^\infty\frac{\beta^{p+1}}{(p+1)!}
(g\mu_{\rm B})^2
 \sum_{l=0}^p \left(\begin{array}{c}p\\l\end{array}\right)
(-J_0)^{p-l}(-J_1)^{l}
\nonumber\\
&& \hspace{0.8cm}\times
\sum_{l_0=0}^{p_{0}} \left(\begin{array}{c} p_{0}\\ l_0 \end{array}\right)
 \sum_{l_1=0}^{p_{1}} \left(\begin{array}{c} p_{1}\\ l_1 \end{array}\right)
\frac{1}{c}
\sum_{\iota,\:\kappa=1}^n 
\sum_{\mu,\:\rho,\:\nu=1}^n 
{\bar u}_{1\iota}
m_\iota^{p_{1}-l_1}V_{\iota\mu}m_\mu^{p_{0}-l_0}V_{\mu\rho}
\tau_p \Delta_\rho^{(p)}
V_{\rho\nu}m_\nu^{l_0}V_{\nu\kappa}m_\kappa^{l_1}
u_{\kappa1}
\nonumber\\
&=&
\frac{(g\mu_{\rm B})^2}{c}
\sum_{p=0}^\infty\frac{\beta^{p+1}}{(p+1)!}
 \sum_{l=0}^p \left(\begin{array}{c}p\\l\end{array}\right)
(-J_0)^{p-l}(-J_1)^{l}
\nonumber\\
&& \hspace{0.8cm}\times
\sum_{\iota,\:\kappa=1}^n 
\sum_{\mu,\:\rho,\:\nu=1}^n 
{\bar u}_{1\iota}
V_{\iota\mu}V_{\mu\rho}
\tau_p \Delta_\rho^{(p)}
V_{\rho\nu}V_{\nu\kappa}
u_{\kappa1}
(m_\mu+m_\nu)^{p-l}(m_\iota+m_\kappa)^{l}
\nonumber\\
&=&
\frac{(g\mu_{\rm B})^2}{c}
\sum_{\iota,\:\kappa=1}^n 
\sum_{\mu,\:\rho,\:\nu=1}^n 
\sum_{p=0}^\infty\frac{\beta^{p+1}}{(p+1)!}
\Big(
(-J_0)(m_\mu+m_\nu)+(-J_1)(m_\iota+m_\kappa)
\Big)^p
{\bar u}_{1\iota}
V_{\iota\mu}V_{\mu\rho}
\tau_p \Delta_\rho^{(p)}
V_{\rho\nu}V_{\nu\kappa}
u_{\kappa1}.
\nonumber\\
\end{eqnarray}
The sum should be classified according to 
$(-J_0)(m_\mu+m_\nu)+(-J_1)(m_\iota+m_\kappa)=J(\iota,\mu,\nu,\kappa)$ and $p$. 
Note that $J(\iota,\mu,\nu,\kappa)^{0}=1$ even if $J(\iota,\mu,\nu,\kappa)=0$. 
Then we find
\beqa
\chi_0
&=&
\frac{(g\mu_{\rm B})^2}{c}
\sum_{\substack{\iota,\:\kappa,\:\mu,\:\nu,\:\rho=1\\ (J(\iota,\mu,\nu,\kappa)=0)}}^n 
\frac{\beta^{1}}{1!}
{\bar u}_{1\iota}
V_{\iota\mu}V_{\mu\rho}
\tau_0 \Delta_\rho^{(0)}
V_{\rho\nu}V_{\nu\kappa}
u_{\kappa 1}
\nonumber\\
&+&
\frac{(g\mu_{\rm B})^2}{c}
\sum_{\substack{\iota,\:\kappa,\:\mu,\:\nu,\:\rho=1\\ (J(\iota,\mu,\nu,\kappa)\neq 0)}}^n 
\sum_{\substack{p=0 \\ (p={\rm even})}}^\infty\frac{\beta^{p+1}}{(p+1)!}
J(\iota,\mu,\nu,\kappa)^p
{\bar u}_{1\iota}
V_{\iota\mu}V_{\mu\rho}
\Delta_{\rho}^{(\rm even)}
V_{\rho\nu}V_{\nu\kappa}
u_{\kappa 1}
\nonumber\\
&+&
\frac{(g\mu_{\rm B})^2}{c}
\sum_{\substack{\iota,\:\kappa,\:\mu,\:\nu,\:\rho=1\\ (J(\iota,\mu,\nu,\kappa)\neq 0)}}^n 
\sum_{\substack{p=1 \\ (p={\rm odd})}}^\infty\frac{\beta^{p+1}}{(p+1)!}
J(\iota,\mu,\nu,\kappa)^p
{\bar u}_{1\iota}
V_{\iota\mu}V_{\mu\rho}
i \Delta_{\rho}^{(\rm odd)}
V_{\rho\nu}V_{\nu\kappa}
u_{\kappa 1},
\eeqa
where $\Delta_{\rho}^{(\rm even)}=\Delta_\rho^{(0)}$ 
and $\Delta_{\rho}^{(\rm odd)}=\Delta_\rho^{(1)}$. 
Finally we obtain 
\beqa
\frac{\chi_{0}}{(g\mu_{\rm B})^2}
&=&
\frac{\beta}{c}
\sum_{\substack{\iota,\:\kappa,\:\mu,\:\nu,\:\rho=1\\ (J(\iota,\mu,\nu,\kappa)=0)}}^n 
{\bar u}_{1\iota}
V_{\iota\mu}V_{\mu\rho}
\Delta_{\rho}^{(0)}
V_{\rho\nu}V_{\nu\kappa}
u_{\kappa 1}
\nonumber\\
&+&
\frac{1}{c}
\sum_{\substack{\iota,\:\kappa,\:\mu,\:\nu,\:\rho=1\\ (J(\iota,\mu,\nu,\kappa)\neq 0)}}^n 
\frac{\sinh \beta J(\iota,\mu,\nu,\kappa)}{J(\iota,\mu,\nu,\kappa)}
{\bar u}_{1\iota}
V_{\iota\mu}V_{\mu\rho}
\Delta_{\rho}^{(\rm even)}
V_{\rho\nu}V_{\nu\kappa}
u_{\kappa 1}
\nonumber\\
&+&
\frac{1}{c}
\sum_{\substack{\iota,\:\kappa,\:\mu,\:\nu,\:\rho=1\\ (J(\iota,\mu,\nu,\kappa)\neq 0)}}^n 
\frac{\cosh \beta J(\iota,\mu,\nu,\kappa)-1}{J(\iota,\mu,\nu,\kappa)}
{\bar u}_{1\iota}
V_{\iota\mu}V_{\mu\rho}
i \Delta_{\rho}^{(\rm odd)}
V_{\rho\nu}V_{\nu\kappa}
u_{\kappa 1}.
\label{susgenform}
\eeqa
Equation (\ref{susgenform}) provides 
the exact zero-field transverse susceptibility 
of the spin-$S$ transverse Ising model (\ref{ham}). 
The result is written in terms of finite summations.  
%there remains no infinite sums or integrals.  

% s=1/2 8/27/2022-1、まとめ9/6/2022-1

In the case of $S=1/2$, 
long and straightforward calculations yield that 
the susceptibility is 
\begin{eqnarray}
\frac{\chi_{0}}{(g\mu_{\rm B})^{2}}
&=&
\frac{\beta}{4c}
\Big[
(x^{2}+\frac{1}{x^{2}})+\frac{y}{R_{-}}(x+\frac{1}{x})
\Big((x-\frac{1}{x})^{2}+\frac{2}{y^{4}}\Big)
\Big]
\nonumber\\
&+&
\frac{1}{J_{0}}\frac{1}{2c}\frac{xy}{R_{-}}
(x^{2}-\frac{1}{x^{2}})
\Big[\frac{R_{-}}{xy}+\frac{1}{x}(x+\frac{1}{x})\Big]
\nonumber\\
&+&
\frac{1}{J_{1}}\frac{1}{2c}\frac{xy}{R_{-}}
\frac{1}{y^{2}}(y^{2}-\frac{1}{y^{2}})
\Big[\frac{R_{-}}{xy}+\frac{1}{x}(x+\frac{1}{x})\Big]
\nonumber\\
&+&
\frac{1}{J_{0}-J_{1}}\frac{1}{2c}\frac{xy}{R_{-}}
\frac{1}{x^{2}}(x^{2}\frac{1}{y^{2}}-\frac{1}{x^{2}}y^{2})
\Big[\frac{y}{x}u_{-}+\frac{1}{y^{2}}\Big]
\nonumber\\
&+&
\frac{1}{J_{0}+J_{1}}\frac{1}{2c}\frac{xy}{R_{-}}
(x^{2}y^{2}-\frac{1}{x^{2}y^{2}})
\Big[xy u_{+}+\frac{1}{y^{2}}\Big],
\label{sus12nnn}
\end{eqnarray}
where 
\begin{eqnarray}
c
&=&
\frac{1}{2}
\Big(y^{4}(x^{4}+\frac{1}{x^{4}})+4(x^{2}+\frac{1}{x^{2}})+4+\frac{2}{y^{4}}\Big)
\nonumber\\
&+&
\frac{1}{2}\frac{y}{R_{-}}(x+\frac{1}{x})
\Big((x^{2}+\frac{1}{x^{2}})+\frac{2}{y^{4}}\Big)
\Big((x-\frac{1}{x})^{2}y^{4}+4\Big)
\nonumber\\
&=&
(\lambda_{1}^{+})^{4},
\label{sus12c}
\end{eqnarray}
and $x=e^{K_{0}/4}$, $y=e^{K_{1}/4}$, 
and $u_{\pm}$ and $R_{-}$ are defined in (\ref{uRpm}). 
Generally in the case of spin $S$, 
we have to find the eigenvectors 
of $n\times n=(2S+1)^{2}\times(2S+1)^{2}$ matrix $V$. 

%------
The first term in (\ref{sus12nnn}) corresponds to the case 
$m_\mu+m_\nu=0$ and $m_\iota+m_\kappa=0$ 
which yields $J(\iota,\mu,\nu,\kappa)=0$ for all $J_{0}$ and $J_{1}$. 
Other $m_{\rho}$'s sometimes yield $J(\iota,\mu,\nu,\kappa)=0$ 
when $J_{0}$ and $J_{1}$ take some special values 
(for example, $J_{1}=-J_{0}$, $m_\mu+m_\nu=2S$, $m_\iota+m_\kappa=2S$ 
result in $J(\iota,\mu,\nu,\kappa)=0$). 
Contributions from these latter cases, if exist, 
can be obtained 
by taking limit $J(\iota,\mu,\nu,\kappa)\to 0$ 
in the remaining four terms 
(thus the susceptibility for $J_{1}=-J_{0}$, for example, 
is simply obtained by taking limit $J_{1}\to -J_{0}$  in (\ref{sus12nnn})). 
%------

When $J_1=0$ or $J_0=0$, 
the system reduces to the nearest-neighbor transverse Ising chain. 
For example in (\ref{sus12nnn}), 
let $J_1\to 0$ then $y\to 1$ and we obtain 
\begin{eqnarray}
\frac{J_{0}\chi_{0}}{(g\mu_{\rm B})^{2}}
=
\frac{K_{0}/2}{\displaystyle (x+\frac{1}{x})^{2}}
+\frac{1}{2}\frac{\displaystyle x-\frac{1}{x}}{\displaystyle x+\frac{1}{x}}
=
\frac{1}{2}\Big(
\frac{K_{0}/4}{\cosh^{2}\frac{1}{4}K_{0}}+\tanh\frac{1}{4}K_{0}
\Big),
\label{sus12}
\end{eqnarray}
which is the transverse susceptibility 
of the nearest-neighbor Ising chain.\cite{60Fisher}\cite{63Fisher}
When $J_0\to 0$ then $x\to 1$ and we again obtain (\ref{sus12}) 
in which $x$ is replaced by $y$, and $J_0$ is replaced by $J_1$.

The susceptibility is invariant 
under the change of sign $J_{0}\mapsto -J_{0}$. 
For example (\ref{sus12nnn}) and (\ref{sus12c}) 
is invariant under $x\mapsto 1/x$. 
This invariance comes from the symmetry of the system under transformations 
$J_{0}\mapsto -J_{0}$ and 
$(\sigma^{x}_{j}, \sigma^{y}_{j}, \sigma^{z}_{j})
\mapsto (\sigma^{x}_{j}, -\sigma^{y}_{j}, -\sigma^{z}_{j})$ 
for odd $j$. 

The transverse susceptibility of the spin $S$ Ising chain 
is shown in Fig.1-Fig.3 
for various values of the nearest-neighbor interaction $J_0$ 
and the next-nearest-neighbor interaction $J_1$. 
The susceptibility is calculated 
using the analytic result (\ref{sus12nnn}) for $S=1/2$, 
and calculated from numerically obtained ${\bar u}_{1\iota}$ and $u_{\kappa 1}$ 
for $S\geq 1$.

%--- 低温極限　10/23/2022-1から
\section{Low-temperature Limit}

Next we physically consider 
the low-temperature behavior of the susceptibility. 
Let us follow the ground-state property of the system. 
Because (\ref{susgenform}) is a finite sum, 
the leading terms in (\ref{susgenform}) 
dominate the susceptibility at $T\to 0$. 
The phase diagram was investigated in 
Refs.\citen{72KatsuraOhminami}-\citen{96Muraoka}, 
and we know the ground state configurations in each phase. 
Let us consider the case $J_{0}>0$, 
because the susceptibility is invariant 
under the change $J_{0}\mapsto -J_{0}$ 
and the result for $J_{0}<0$ and $J_{1}$ 
is identical with that for $|J_{0}|$ and $J_{1}$.

When $J_{1}>-J_{0}/2$, 
the ground state is ferromagnetic: 
$s_{j}=S$ for all $j$, or $s_{j}=-S$ for all $j$. 
The leading terms in (\ref{cgenform}) are 
\begin{eqnarray}
{\bar u_{11}}(e^{S^{2}K_{0}+S^{2}K_{1}})^{4}u_{11}
+{\bar u_{1n}}(e^{(-S)^{2}K_{0}+(-S)^{2}K_{1}})^{4}u_{n1}.
\end{eqnarray}
Note that ${\bar u_{1\iota}}u_{\rho 1}$ 
is the exactly renormalized Boltzmann weight 
with fixed spin configurations $\iota$ and $\kappa$, 
and satisfies ${\bar u_{1\iota}}u_{\rho 1}>0$.  
Because $U^{-1}U=E$, 
we find that $\sum_{\iota=1}^{n}{\bar u_{1\iota}}u_{\iota 1}=1$, 
and thus all the ${\bar u_{1\iota}}u_{\kappa 1}$ are finite when $\iota=\kappa$. 
The leading terms in (\ref{susgenform}) 
are found in $(1,1)$ and $(n,n)$-elements of $V^{2}\tau_{p}D^{(p)}V^{2}$, 
with the condition 
$m_{\iota}+m_{\kappa}=m_{\mu}+m_{\nu}=2S$ and 
$m_{\iota}+m_{\kappa}=m_{\mu}+m_{\nu}=-2S$, respectively. 
They come from the ground-state configurations 
together with the first excitations: 
\begin{eqnarray}
&&
|S\rangle_{j-2}|S\rangle_{j-1}|S\rangle_{j}|S\rangle_{j+1}|S\rangle_{j+2}|S\rangle_{j+3}, 
\nonumber\\
&&
|S\rangle_{j-2}|S\rangle_{j-1}|S-1\rangle_{j}|S\rangle_{j+1}|S\rangle_{j+2}|S\rangle_{j+3} 
\end{eqnarray}
and 
\begin{eqnarray}
&&
|-S\rangle_{j-2}|-S\rangle_{j-1}|-S\rangle_{j}|-S\rangle_{j+1}|-S\rangle_{j+2}|-S\rangle_{j+3},
\nonumber\\
&&
|-S\rangle_{j-2}|-S\rangle_{j-1}|-S+1\rangle_{j}|-S\rangle_{j+1}|-S\rangle_{j+2}|-S\rangle_{j+3},
\end{eqnarray}
i.e. obtained from the terms 
\begin{eqnarray}
\frac{1}{c}&\Big[&
\frac{\sinh(-2S\beta(J_{0}+J_{1}))}{-2S(J_{0}+J_{1})}
\nonumber\\
&&
\hspace{1.2cm}\times
\Big(
\Delta_{1}^{({\rm even})}
(e^{S^{2}K_{0}+S^{2}K_{1}})^{4}
+\Delta_{2}^{({\rm even})}
(e^{S^{2}K_{0}+S^{2}K_{1}})^{2}(e^{S(S-1)K_{0}+S(S-1)K_{1}})^{2}
\Big)
{\bar u_{11}}u_{11}
\nonumber\\
&&+
\frac{\sinh(2S\beta(J_{0}+J_{1}))}{2S(J_{0}+J_{1})}
\nonumber\\
&&
\hspace{1.2cm}\times
\Big(
\Delta_{n}^{({\rm even})}
(e^{(-S)^{2}K_{0}+(-S)^{2}K_{1}})^{4}
+\Delta_{n-1}^{({\rm even})}
(e^{(-S)^{2}K_{0}+(-S)^{2}K_{1}})^{2}(e^{(-S)(-S+1)K_{0}+(-S)(-S+1)K_{1}})^{2}
\Big)
{\bar u_{1n}}u_{n1}
\nonumber\\
&&+
\frac{\cosh(-2S\beta(J_{0}+J_{1}))-1}{-2S(J_{0}+J_{1})}
\nonumber\\
&&
\hspace{1.2cm}\times
\Big(
i\Delta_{1}^{({\rm odd})}
(e^{S^{2}K_{0}+S^{2}K_{1}})^{4}
+i\Delta_{2}^{({\rm odd})}
(e^{S^{2}K_{0}+S^{2}K_{1}})^{2}(e^{S(S-1)K_{0}+S(S-1)K_{1}})^{2}
\Big)
{\bar u_{11}}u_{11}
\nonumber\\
&&+
\frac{\cosh(2S\beta(J_{0}+J_{1}))-1}{2S(J_{0}+J_{1})}
\nonumber\\
&&
\hspace{1.2cm}\times
\Big(
i\Delta_{n}^{({\rm odd})}
(e^{(-S)^{2}K_{0}+(-S)^{2}K_{1}})^{4}
+i\Delta_{n-1}^{({\rm odd})}
(e^{(-S)^{2}K_{0}+(-S)^{2}K_{1}})^{2}(e^{(-S)(-S+1)K_{0}+(-S)(-S+1)K_{1}})^{2}
\Big)
{\bar u_{1n}}u_{n1}
\Big],
\nonumber\\
\end{eqnarray}
where 
$\Delta_{1}^{({\rm even})}=\Delta_{n}^{({\rm even})}
=\frac{1}{2}(S(S+1)-S^{2})=\frac{1}{2}S$, 
$\Delta_{2}^{({\rm even})}=\Delta_{n-1}^{({\rm even})}
=\frac{1}{2}(S(S+1)-(S-1)^{2})=\frac{1}{2}(3S-1)$, 
$i\Delta_{1}^{({\rm odd})}=\frac{1}{2}S$, 
$i\Delta_{2}^{({\rm odd})}=\frac{1}{2}(S-1)$, 
$i\Delta_{n-1}^{({\rm odd})}=-\frac{1}{2}(S-1)$, 
and 
$i\Delta_{n}^{({\rm odd})}=-\frac{1}{2}S$. 
Contributions from the first sum in (\ref{susgenform}) vanish 
at $T\to 0$. 
Short calculations yield that the susceptibility at low temperatures behaves 
\begin{eqnarray}
\frac{\chi_{0}}{(g\mu_{\rm B})^{2}}
\to
\frac{1}{2}\frac{1}{J_{0}+J_{1}}
\hspace{0.6cm}
(T\to 0),
\end{eqnarray}
which is independent of $S$. 

When $J_{1}<-J_{0}/2$, 
the ground state configurations are obtained by iterations of 
$|S\rangle|S\rangle|-S\rangle|-S\rangle$.
\cite{96Muraoka}\cite{72MoritaHoriguchi} 
Specifically from the configurations 
\begin{eqnarray}
&&
|S\rangle_{j-2}|S\rangle_{j-1}|-S\rangle_{j}|-S\rangle_{j+1}|S\rangle_{j+2}|S\rangle_{j+3}, 
\hspace{0.9cm}
|S\rangle_{j-2}|-S\rangle_{j-1}|-S\rangle_{j}|S\rangle_{j+1}|S\rangle_{j+2}|-S\rangle_{j+3}, 
\nonumber\\
&&
|-S\rangle_{j-2}|-S\rangle_{j-1}|S\rangle_{j}|S\rangle_{j+1}|-S\rangle_{j+2}|-S\rangle_{j+3}, 
\hspace{0.5cm}
|-S\rangle_{j-2}|S\rangle_{j-1}|S\rangle_{j}|-S\rangle_{j+1}|-S\rangle_{j+2}|S\rangle_{j+3}, 
\nonumber\\
\end{eqnarray}
and their excitations, 
the leading terms of (\ref{susgenform}) are generated 
in the diagonal elements of $V^{2}\tau_{p}D^{(p)}V^{2}$, 
and again straightforwrd calculations yield that 
the low-temperature limit of the susceptibility is 
\begin{eqnarray}
\frac{\chi_{0}}{(g\mu_{\rm B})^{2}}
\to
\frac{-1}{2}\frac{1}{J_{1}}
=
\frac{1}{2}\frac{1}{|J_{1}|}
\hspace{0.6cm}
(T\to 0).
\end{eqnarray}
Note that the nearest-neighbor interactions cancel with each other 
and $J_{0}$ does not appear in the ground state configurations. 

When $J_{1}=-J_{0}/2$, 
configurations obtained by arbitrary iterations of 
$|S\rangle|S\rangle$ and $|-S\rangle|-S\rangle$ 
belong to the ground state, 
e.g.  
configurations 
$|S\rangle|S\rangle|S\rangle|S\rangle|S\rangle|S\rangle\cdots$, 
and 
$|S\rangle|S\rangle|-S\rangle|-S\rangle|S\rangle|S\rangle\cdots$, 
and 
$|S\rangle|S\rangle|S\rangle|S\rangle|-S\rangle|-S\rangle\cdots$ 
provide the ground state energy. 
The ground state is therefore highly degenerate, 
and for example the configurations 
\begin{eqnarray}
|S\rangle_{j-2}|S\rangle_{j-1}|-S\rangle_{j}|-S\rangle_{j+1}|-S\rangle_{j+2}|-S\rangle_{j+3}, 
\hspace{0.6cm}
|S\rangle_{j-2}|S\rangle_{j-1}|S\rangle_{j}|-S\rangle_{j+1}|-S\rangle_{j+2}|-S\rangle_{j+3}, 
\end{eqnarray}
generate some of the leading terms. 
They satisfy $m_{\iota}+m_{\kappa}=0$ and $m_{\mu}+m_{\nu}=0$, 
and hence the first term in (\ref{susgenform}) remains even when $T\to 0$, 
and we find the divergence coming from $\beta$ at low temperatures.

\begin{acknowledgment}
%\acknowledgment
This work was supported by JSPS KAKENHI Grant No. JP19K03668.
\end{acknowledgment}

%------ Figures

\begin{figure}
\includegraphics[width=15.0cm]{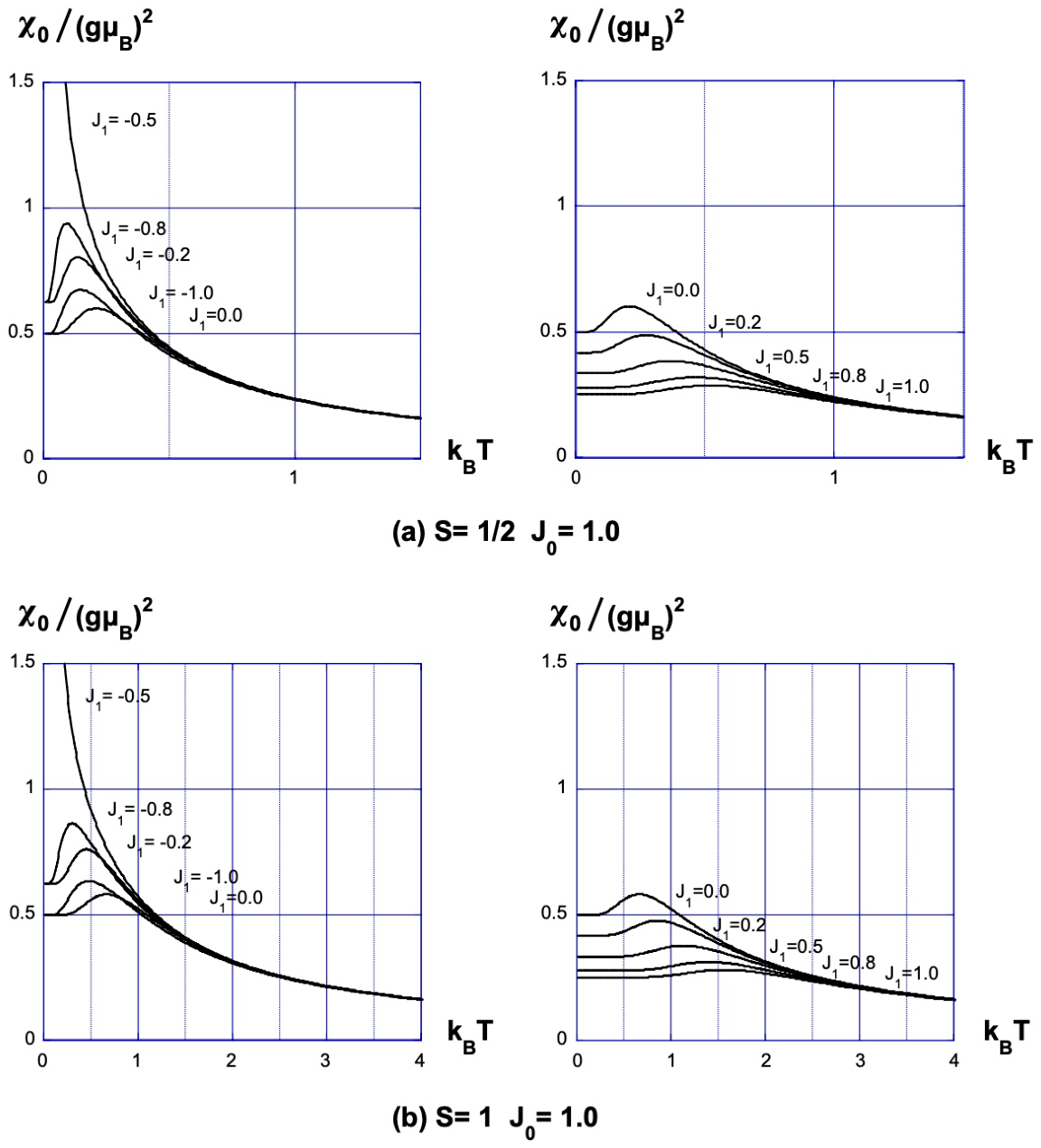}
\end{figure}

\begin{figure}
\includegraphics[width=15.0cm]{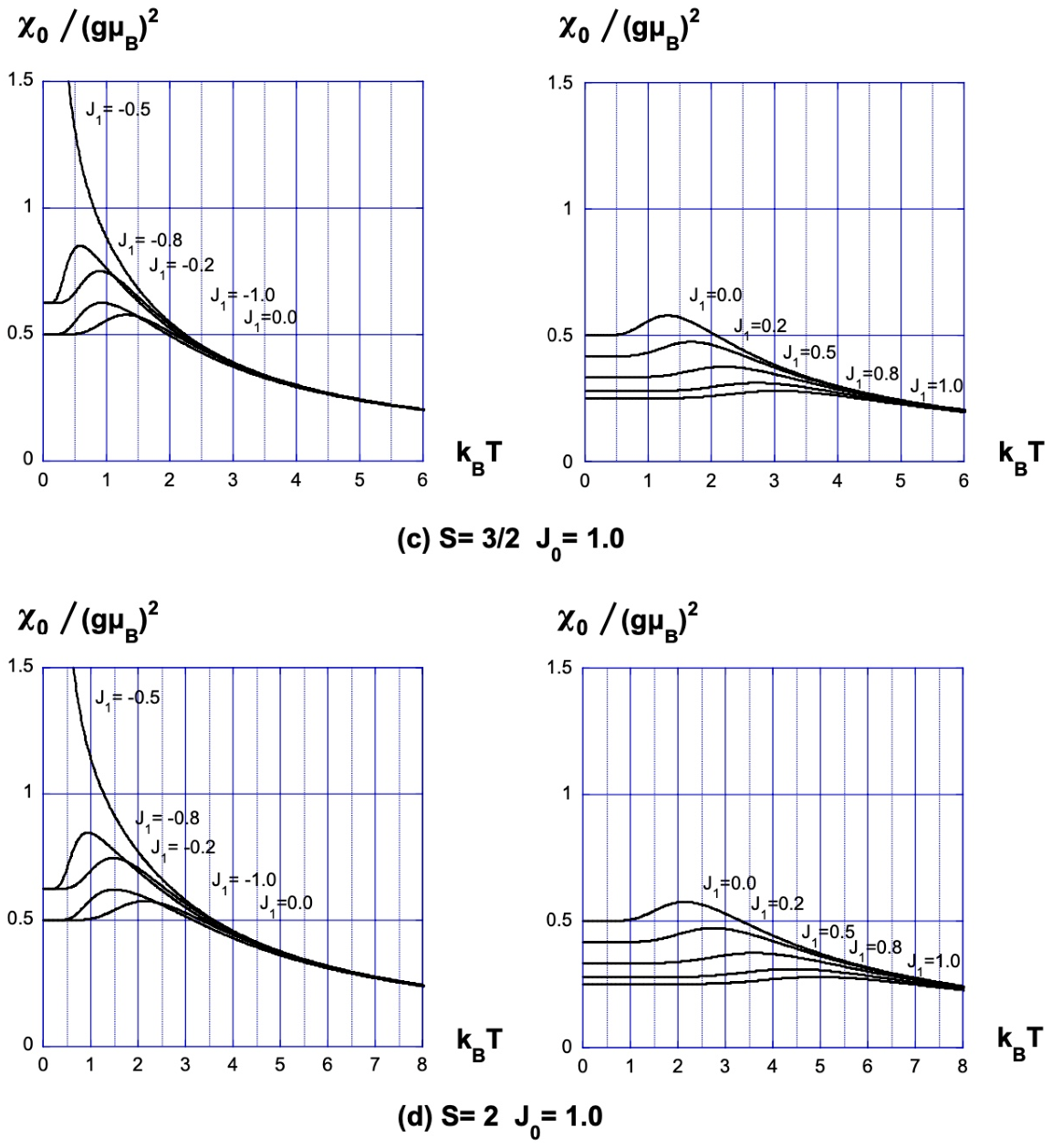}
\caption{
The transverse susceptibility of the spin $S$ Ising chain 
with the nearest-neighbor-interaction $J_0=1.0$, 
and with various values of the next-nearest-neighbor interaction $J_1$, 
where 
(a) $S=1/2$,  
(b) $S=1$,  
(c) $S=3/2$,  
and 
(d) $S=2$.  
}
\end{figure}

\begin{figure}
\includegraphics[width=15.0cm]{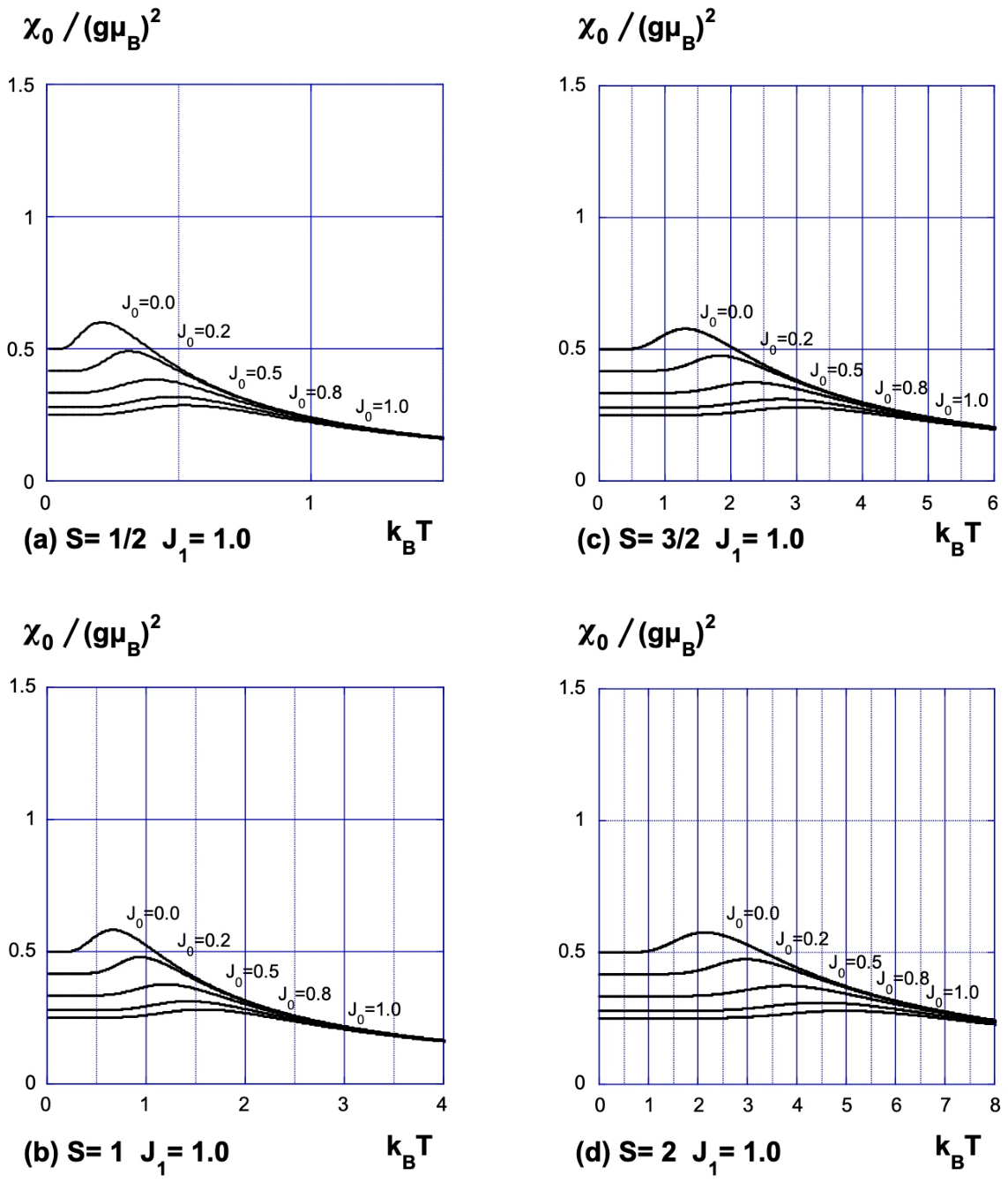}
\caption{
The transverse susceptibility of the spin $S$ Ising chain 
with the next-nearest-neighbor-interaction $J_1=1.0$, 
and with various values of the nearest-neighbor interaction $J_0>0$, 
where 
(a) $S=1/2$,  
(b) $S=1$,  
(c) $S=3/2$,  
and 
(d) $S=2$.  
}
\end{figure}

\begin{figure}
\includegraphics[width=15.0cm]{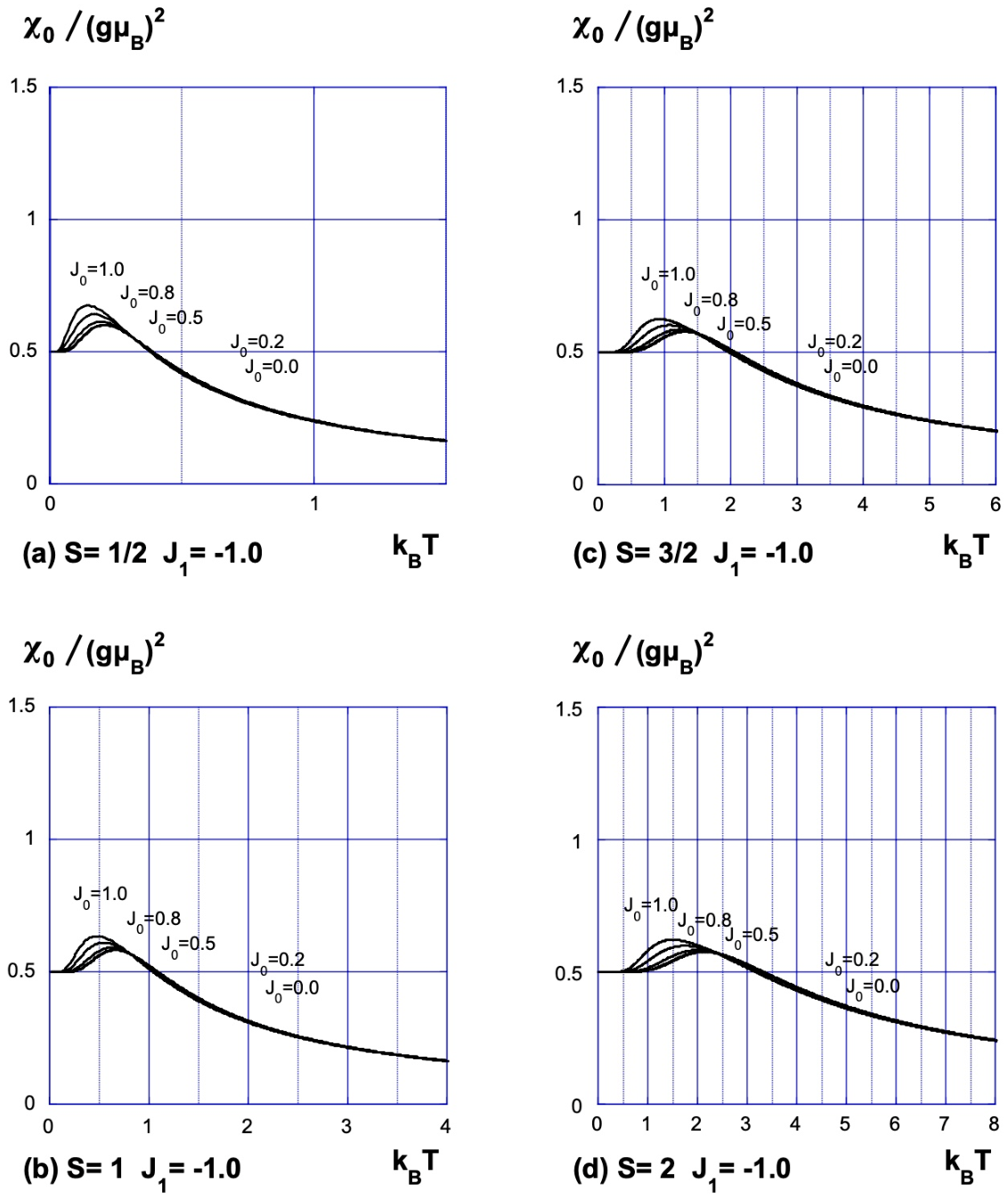}
\caption{
The transverse susceptibility of the spin $S$ Ising chain 
with the next-nearest-neighbor-interaction $J_1=-1.0$, 
and with various values of the nearest-neighbor interaction $J_0>0$, 
where 
(a) $S=1/2$,  
(b) $S=1$,  
(c) $S=3/2$,  
and 
(d) $S=2$. 
The curves for $J_{0}=0.0$ and for $J_{0}=0.2$ are very close to each other. 
The curve for $J_{0}=1.0$ is larger than the other four at low temperatures, 
and is smaller than the other four at high temperatures. 
}
\end{figure}

%------ references

\end{document}